%% file: main.tex
\documentclass[aps,prl,reprint,nofootinbib,amsmath,amssymb,floatfix]{revtex4-2}

\usepackage[T1]{fontenc}
\usepackage{graphicx}
\usepackage[dvipsnames]{xcolor}
\usepackage[hidelinks]{hyperref}
\usepackage{orcidlink}
\usepackage{tikz}
\usepackage[normalem]{ulem}
\usetikzlibrary{arrows.meta}
\usepackage{bm}
\usepackage{mathrsfs}
\graphicspath{{./plots/}}
\newcommand{\ii}{\mathrm{i}}
\newcommand{\Th}{\text{\TH}}
\newcommand{\scri}{\mathscr{I}}
\begin{document}

\title{Cubic Quasinormal Modes in the Black-Hole Ringdown}

\author{Neev Khera \orcidlink{0000-0003-3515-2859}}
\email{nkhera@tsinghua.edu.cn}
\affiliation{Department of Astronomy, Tsinghua University, Beijing 100084, China}

\author{Huan Yang \orcidlink{0000-0002-9965-3030}}
\email{hyangdoa@tsinghua.edu.cn}
\affiliation{Department of Astronomy, Tsinghua University, Beijing 100084, China}

\begin{abstract}
We formulate and numerically solve the nonlinear Teukolsky equations through third order, obtaining a first-principles calculation of cubic quasinormal-mode amplitudes for Kerr black holes. Our approach combines lower-order metric reconstruction in split light-cone gauges with a hyperboloidal, compactified formulation solved on complex radial contours.
Spin-dependent complex coupling coefficients fix the cubic amplitudes from products of their linear parent amplitudes.
We apply these predictions to a nonprecessing numerical-relativity waveform, including cubic products of the $(2,2)$ fundamental mode, its first overtone, and their conjugates. The predicted cubic terms improve agreement over selected early-ringdown fitting windows and shift inferred linear-overtone amplitudes without adding fitting parameters. $3M$ after the peak of $|h_{22}|$, the summed cubic contribution reaches approximately $38\%$ of the total $h_{22}$ signal; $M$ is the initial total mass. These results identify an appreciable cubic response in early ringdown.
\end{abstract}

\maketitle

\paragraph{Introduction.---}

The gravitational waves emitted after a binary black-hole merger probe the relaxation of the remnant toward equilibrium. Their quasinormal-mode (QNM) frequencies and damping times are fixed, in linear perturbation theory, by the remnant's mass and spin~\cite{Teukolsky1973,Leaver1985,BertiCardosoWill2006}. Measuring several modes tests whether their complex frequencies are consistent with a single Kerr spacetime, the basis of black-hole spectroscopy \cite{Berti:2018vdi,Berti:2025hly,BertiCardosoWill2006}. Observations of GW250114 have enabled increasingly stringent tests of this structure~\cite{LVK2026Spectroscopy}.

Whereas the QNM frequencies characterize the remnant, their excitation amplitudes retain information about the progenitor and merger dynamics \cite{Hughes:2019zmt,Becker:2025zzw,Crescimbeni:2025kxi,Cheung:2023vki,London:2014cma,London:2018gaq,KamaretsosHannamSathyaprakash2012,SuEtAl2026Kerr}. Ringdown thus offers complementary probes of the final black hole and the dynamical process that formed it. The postmerger waveform, however, is not composed solely of linear QNMs: it also contains direct radiation~\cite{Leaver1986,Andersson1997,SuEtAl2026Schwarzschild,SuEtAl2026Kerr} and late-time tails~\cite{Price1972I,Price1972II,GundlachPricePullin1994,ChingEtAl1995,Barack2000,DeAmicisEtAl2025}, as well as contributions from nonlinear interactions~\cite{Zlochower:2003yh,IokaNakano2007,OkuzumiIokaSakagami2008,LagosHui2023,MitmanEtAl2023,CheungEtAl2023}. Interpreting the observed ringing components therefore requires understanding how these different contributions enter the signal, both when testing the Kerr spectrum and when inferring the physical excitation amplitudes.

Nonlinear perturbations can contribute appreciably to the early ringdown, making their inclusion important for accurate waveform modeling \cite{MitmanEtAl2023,CheungEtAl2023,Ripley:2020xby}. Beyond this practical role, their mode-coupling amplitudes and phases probe gravitational self-interactions, providing tests of the nonlinear structure of Einstein's equations beyond the linear QNM spectrum \cite{LagosHui2023,MaYang2024,KheraMaYang2025}. Substantial progress has been made at quadratic order: quadratic QNMs have been identified in numerical-relativity waveforms~\cite{MitmanEtAl2023,CheungEtAl2023,Cheung:2023vki} and horizon dynamics~\cite{Khera:2023oyf}, and their excitation amplitudes and coupling coefficients have been calculated using second-order perturbation theory~\cite{NakanoIoka2007,LagosHui2023,MaYang2024,Bucciotti:2024zyp,Bucciotti:2024jrv,Bourg:2024jme,KheraMaYang2025}. Evidence for quadratic QNMs has also been reported in GW250114 \cite{Wang:2026rev}. Beyond quadratic order, studies of absorption-induced backreaction have shown that changes in the black hole’s mass and spin can produce significant third-order corrections to mode excitation and waveform evolution \cite{Sberna:2021eui,Zhu:2024dyl,May:2024rrg}. More recently, cubic QNM contributions have been identified in numerical-relativity ringdowns \cite{Dyer:2025hdt}.
Cubic products of $(2,2)$ modes, with one factor conjugated, can contribute back to the dominant $(2,2)$ multipole. Their proximity in frequency to linear overtones allows linear-only fits to absorb cubic response, biasing the inferred linear amplitudes. Predicting cubic couplings is therefore important for interpreting these amplitudes.

In this Letter, we construct and numerically solve the nonlinear Teukolsky equations through third order, establishing the first framework for quantitatively predicting cubic QNM amplitudes of Kerr black holes. Our approach combines metric reconstruction in complementary light-cone gauges with a numerical solution on complex radial contours; the full formulation and implementation will appear separately~\cite{KheraYangMethods2026}. By determining cubic amplitudes from their linear parents, the theory enables a quantitative search for cubic modes in numerical-relativity waveforms without introducing additional fitting parameters. As a first application, we analyze a nonprecessing binary-black-hole waveform and find that cubic modes contribute appreciably to the dominant $(2,2)$ strain multipole during early ringdown. These results demonstrate the importance of cubic response even in the strongest component of the signal, providing a quantitative basis for its incorporation into ringdown modeling and spectroscopy.

\paragraph{Third-order perturbation.---}
Write the metric as
$g_{ab}=g^{(0)}_{ab}+\epsilon h^{(1)}_{ab}
+\epsilon^2 h^{(2)}_{ab}+\epsilon^3 h^{(3)}_{ab}+\cdots$, where $g^{(0)}_{ab}$
is the Kerr background with remnant mass $M_f$ and dimensionless spin $\chi_f$, $h^{(k)}_{ab}$ is the order-$k$ perturbation, and $\epsilon$ is a
bookkeeping parameter. Let $G^{[k]}_{ab}$ denote the order-$k$ variation of the Einstein tensor with respect to the metric.
Expanding Einstein's equations to order $k$ gives
\begin{align}
  G^{[1]}_{ab}[h^{(k)}]
  &=
-\sum_{n=2}^{k}\frac{1}{n!}
\sum_{\substack{k_1+\cdots+k_n=k}}
G^{[n]}_{ab}
\!\left[
h^{(k_1)},\ldots,h^{(k_n)}\,,
\right]
\nonumber \\
  &= 8\pi T_{ab}^{(k),\mathrm{eff}}\,,
\label{eq:order-k-source}
\end{align}
where $k_i\geq 1$. Thus, $h_{ab}^{(k)}$ satisfies linearized Einstein's equations with an effective stress-energy tensor $T_{ab}^{(k),\mathrm{eff}}$ determined by lower-order perturbations. The reduced Weyl scalars $\Psi^{(k)}_4[h_{ab}^{(k)}]$ and $\Psi^{(k)}_0[h_{ab}^{(k)}]$---defined as the parts of the Weyl scalar perturbation linear in $h_{ab}^{(k)}$---satisfy Teukolsky equations~\cite{Spiers:2023cip,Green:2019nam,Campanelli:1998jv}:
\begin{equation}
     \label{eq:teukolsky-eq}
    \mathcal{O^\prime} \Psi_4^{(k)} = S_4^{(k)} \quad \mathrm{and} \quad \mathcal{O} \Psi_0^{(k)} = S_0^{(k)}\,,
\end{equation}
where $\mathcal{O}^\prime$ and $\mathcal{O}$ are the Teukolsky operators and $S_4$ and $S_0$ are their sources, computed from the effective stress-energy tensor $T_{ab}^{(k),\mathrm{eff}}$~\cite{Pound:2021qin, Teukolsky1973}. The asymptotic gravitational wave amplitude at order-$k$ is extracted from $\Psi_4^{(k)}$. To obtain the order-$(k+1)$ amplitudes, we must reconstruct the metric $h_{ab}^{(k)}$ from $\Psi_0^{(k)}$ and $\Psi_4^{(k)}$.
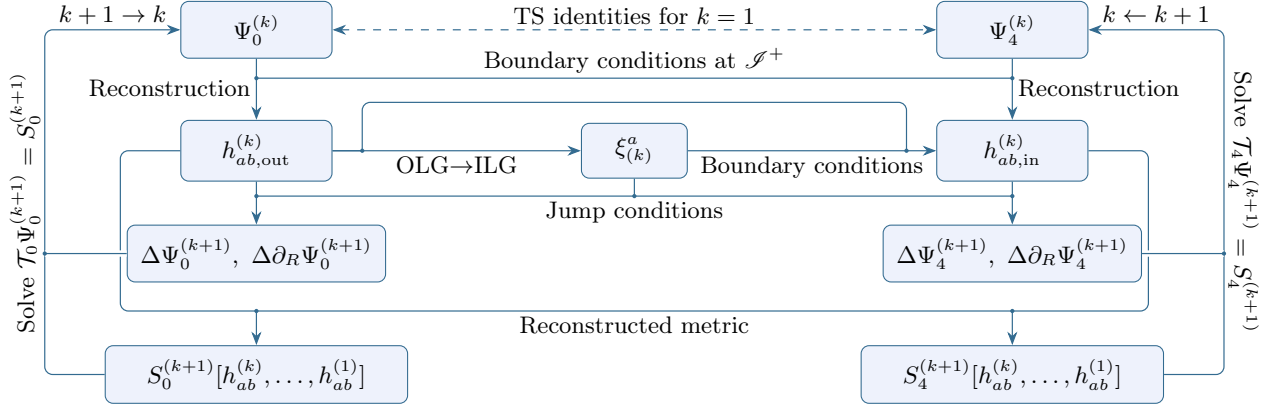
\begin{figure*}
    \centering
    \definecolor{flowblue}{HTML}{336b90}
    \definecolor{flowink}{HTML}{336b90}
    \definecolor{flowfill}{HTML}{eef2fc}
    \input{flowchart.tikz}
    \caption{Iterative construction of nonlinear perturbations.
    At order $k$, the reduced Weyl fields determine the metric in
    outer and inner light-cone gauges. The reconstructed metrics and
    their gauge transformation supply the sources and jump conditions
    for the order-$(k+1)$ Teukolsky equations. Iterating through $k=2$
    yields the third-order Weyl fields. Here $R=1/r$ is the
    compactified radial coordinate.}
    \label{fig:flowchart}
\end{figure*}

In this Letter, we implement this procedure for QNMs. Label a Kerr QNM by $q=(\ell,m,n,p)$, where $(\ell,m)$ are its spheroidal
angular indices, $n$ its overtone, and $p=\pm$ its frequency branch \cite{MaYang2024,KheraMaYang2025}.
Let $\omega_q$ be its complex frequency and $A_q$ its spherical strain
amplitude in the $(\ell_q,m_q)$ projection. Reality of the metric requires a conjugated contribution
$\bar q$, with $\omega_{\bar q}=-\bar\omega_q$,
$m_{\bar q}=-m_q$, and $A_{\bar q}=\bar A_q$.

At order $k$, let $Q=q_1\cdots q_k$ label a product of linear QNM
contributions, each barred or unbarred. Its perturbative order is $|Q|=k$. Its frequency and azimuthal
number are $\omega_Q=\sum_{i=1}^k\omega_{q_i}$ and
$m_Q=\sum_{i=1}^k m_{q_i}$, and its amplitude is proportional to
$\prod_{i=1}^k A_{q_i}$. We solve the nonlinear equations one product $Q$ at a time.
For each product $Q$, we denote the corresponding reduced Weyl fields by
$\Psi_0^Q$ and $\Psi_4^Q$.
They satisfy Teukolsky equations with sources $S_0^Q$ an $S_4^Q$, the $Q$ components of the effective sources in Eq.~\eqref{eq:teukolsky-eq}.

Computing the third-order source requires reconstructing the sourced
second-order metric as well as the vacuum first-order metric~\cite{Loutrel:2020wbw,Price:2006ke}.
Reconstruction in a single light-cone gauge can become singular
where its defining null congruence develops caustics.
We therefore use the ingoing light-cone gauge (ILG) and outgoing
light-cone gauge (OLG)~\cite{Preston:2006ze} in two radial regions:
\begin{equation}
  h_{ab}^{\mathrm{in}}n^b=0\quad(r<r_c),\qquad
  h_{ab}^{\mathrm{out}}\ell^b=0\quad(r>r_c).
\end{equation}
Here $\ell^a$ and $n^a$ are the outgoing and ingoing principal null directions,
and $r_c$ is the radius at which we switch gauges.
The split avoids the caustic-prone region of each gauge:
near the horizon for the $\ell^a$-adapted gauge and near infinity
for the $n^a$-adapted gauge. A gauge generator
$\xi^a=\sum_{j\geq1}\epsilon^j\xi^a_{(j)}$, chosen to vanish at
$r=r_c$, relates the two reconstructions.

The outgoing condition requires setting the coefficient of an exponentially subdominant incoming solution to zero at null infinity. This is poorly conditioned on the real radial domain, especially for purely imaginary frequencies, motivating the complex-contour formulation described below.

At each order, we first solve the sourced Teukolsky equations for
$\Psi_0^Q$ and $\Psi_4^Q$ on hyperboloidal slices using spectral methods~\cite{PanossoMacedo:2019npm,Ripley:2022ypi}. At linear order, the Teukolsky-Starobinsky identities fix the relative normalization of the Weyl fields; at higher orders, their normalization is fixed by the sources and boundary conditions. Beyond linear order, these reduced fields depend
on the lower-order gauge, so their values and radial derivatives obey
jump conditions at $r_c$ determined by the lower-order gauge transformation.
We reconstruct the outer metric by Newman--Penrose transport from
$\Psi_0^Q$, using $\Psi_4^Q$ and the asymptotic Bianchi identities to fix
boundary data at null infinity~\cite{NewmanPenrose1962,GerochHeldPenrose1973}, see Appendix~\ref{sec:supp-transport} for details. Transforming that metric at $r_c$ supplies
data for the inner reconstruction from $\Psi_4^Q$. Applying this procedure through second
order supplies the metrics needed to compute the third-order Weyl fields.
We summarize this procedure in Figure~\ref{fig:flowchart}.

We approach null infinity along an arc in the complex radial plane. Conjugated channels require the fields on the reflected arc as well, so we solve on both contours. Outgoing boundary conditions must be imposed in the same Stokes sector as the real axis; the sign of $\operatorname{Re}\omega$ selects the appropriate arc. For purely imaginary frequencies ($m=0$), the real axis lies on a Stokes line, and we average the two contour prescriptions. Further details are given in Appendix~\ref{sec:supp-contour}.

Once $\Psi_4^Q$ is obtained, its spherical-harmonic projection gives the
single-channel strain amplitude $A^Q_{\ell m}$. Other channels can
contribute at the same frequency.
Define $q^\star=(\ell,-m,n,-p)$ to be the reflected QNM of $q$. Since
$\omega_{q^\star}=-\bar\omega_q$, its conjugated contribution
$\overline{q^\star}$ has the same frequency and azimuthal number as $q$.
For a cubic product $Q=q_1q_2q_3$, each parent factor can take either
form, yielding up to eight channels before repeated parents are
identified. The channel
in which all three parents are conjugated has analytically vanishing
outgoing strain amplitude in this construction, so it does not contribute
to the channel sum. Let $\langle Q\rangle$ denote the channels with frequency
$\omega_Q$, and define their summed amplitude by
$A^{\langle Q\rangle}_{\ell m}
\equiv\sum_{Q'\in\langle Q\rangle}A^{Q'}_{\ell m}$.
Symmetry-equivalent channel combinations are counted only once.
For waveforms from nonprecessing binaries, reflection symmetry gives
$A_{q^\star}=(-1)^{\ell_q}\bar A_q$, where $\ell_q$ is the
spheroidal degree of $q$. The sum then factorizes:
\begin{equation}
  A^{\langle Q\rangle}_{\ell m}
  =Z^Q_{\ell m} A_{q_1}A_{q_2}A_{q_3}\,,
  \label{eq:cubic-ratio}
\end{equation}
The complex dimensionless coupling coefficient $M_f^2 Z^Q_{\ell m}$ encodes the cubic response
of the black hole and depends only on its spin. 

\paragraph{Cubic quasinormal modes.---}
To specify the parent amplitudes in Eq.~\eqref{eq:cubic-ratio}, write the
linear strain as
\begin{align}
  h^{(1)}(t,\theta,\phi)
  = \sum_{q} \mathcal{A}^{S}_q S_q(\theta,\phi) e^{- i\omega_q t}\,,
  \label{eq:linear-model}
\end{align}
Here $t=0$ is the
peak of $|h_{22}|$. The quantities $\omega_q$ and
$\mathcal A_q^{S}$ are the mode's complex frequency and spheroidal strain
amplitude at that reference time. With the remnant's spin
parameter $a=M_f \chi_f$, our angular convention is
\begin{equation}
  S_q(\theta,\phi)
  ={}_{-2}S_{\ell m}(\theta;-a\omega_q)e^{im\phi}
  =\sum_{\ell^\prime} C^q_{\ell^\prime}\,{}_{-2}Y_{\ell^\prime m}(\theta,\phi),
\end{equation}
where ${}_{-2}Y_{\ell^\prime m}$ is a spin-weight $-2$ spherical harmonic and
$C^q_{\ell^\prime}$ is the spheroidal--spherical mixing coefficient. We
use the corresponding spherical amplitude
$A_q=C^q_{\ell_q}\mathcal A_q^S$.

For convenience of notation, we write $I_n\equiv(2,2,n,+)$ for the co-rotating $(2,2)$ QNM family, with
$\bar I_n$ denoting its conjugated contribution. We select cubic products
built from $I_0$, $I_1$, and their conjugates, with resulting azimuthal index $m=2$.
The six products are
$I_0^2\bar I_0$,
$I_0^2\bar I_1$,
$I_0I_1\bar I_0$,
$I_0I_1\bar I_1$,
$I_1^2\bar I_0$, and
$I_1^2\bar I_1$.
The imaginary parts of these cubic frequencies cluster around those of
successive linear $(2,2)$ overtones. Since
$\gamma=-\operatorname{Im}\omega$ sets the decay timescale, cubic and linear
modes within each cluster persist over comparable intervals.
We therefore truncate QNM models by damping-rate group,
retaining cubic modes alongside their linear-overtone counterparts in each
included cluster. Let $N$ be the highest included $(2,2)$ overtone.
We assign each cubic product to the $I_n$ nearest in damping rate and
include it when $n\leq N$.
The $N=2,3,4$ models thus contain three, four, and five linear $(2,2)$ modes
and three, five, and six cubic products, respectively.

\begin{figure}[t]
  \centering
  \includegraphics{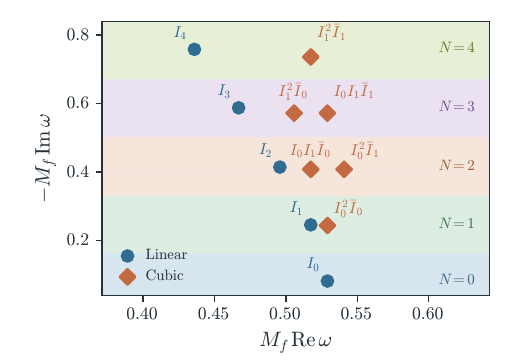}
  \caption{Linear and cubic frequencies at the SXS:BBH:0305 remnant
  parameters. Here $I_n=(2,2,n,+)$; a bar denotes a conjugated contribution.
  Light bands label the damping-rate groups $N=0,\ldots,4$; cubic frequencies
  are labeled by their products and grouped with the $I_N$ nearest in
  damping rate $-\operatorname{Im}\omega$.}
  \label{fig:frequencies}
\end{figure}

The frequency separations vary with spin, and closer linear--cubic pairs
may require a treatment of resonant response. Figure~\ref{fig:frequencies}
illustrates the damping-rate groups as bands at the SXS:BBH:0305 remnant
spin, $\chi_f=0.692$, close to that inferred for
GW150914~\cite{LIGOScientific:2016vlm}.

\begin{figure}[t]
  \centering
  \includegraphics{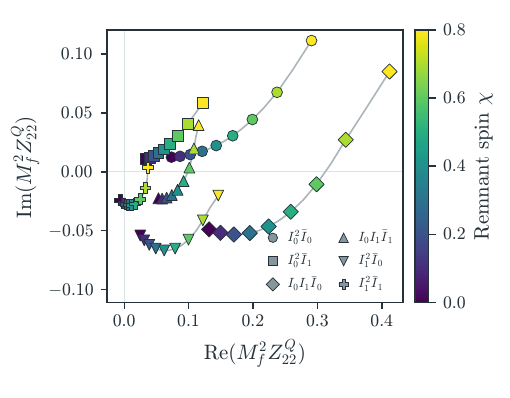}
  \caption{Nonprecessing cubic coupling coefficients $Z^Q_{22}$ for the six
  products in Fig.~\ref{fig:frequencies}. Lines connect samples over
  $0\leq\chi\leq0.8$ in steps of $0.05$; markers are shown every $0.1$.
  Marker shape identifies the product and color gives $\chi$.}
  \label{fig:spin-ratios}
\end{figure}

Figure~\ref{fig:spin-ratios} shows the six complex coupling coefficients
$Z^Q_{22}$ over $0\leq\chi\leq0.8$, where $\chi=J_f/M_f^2$ is the
dimensionless remnant spin. Their magnitudes are comparable, so the
relative strengths of these cubic contributions are governed primarily
by the linear parent amplitudes and their summed damping rates.

\paragraph{Search in numerical waveforms.---}

We analyze the $h_{22}$, $h_{32}$, and $h_{42}$ harmonics of SXS:BBH:0305~\cite{Boyle:2019kee},
fitting each interval $[t_0,60M]$ independently. Every fitted amplitude
refers to its value at the peak of $|h_{22}|$, independently of $t_0$.
The linear basis includes spheroidal $(\ell,2)$ overtones through $n=N$
for $\ell=2,3,4$, with each amplitude shared across the three spherical
harmonics through the mixing coefficients $C^q_{\ell'}$~\cite{MaganaZertuche:2021syq,Li:2021wgz}.

In terms of these spherical amplitudes, the joint model is
\begin{equation}
\begin{aligned}
  h_{\ell 2}^{\mathrm{model}}(t)
  ={}&\sum_q \frac{C_\ell^q}{C_{\ell_q}^q}A_q e^{-i\omega_q t}
  \\
  &+\sum_{Q=q_iq_j\bar q_k}
  Z_{\ell 2}^Q A_{q_i}A_{q_j}\bar A_{q_k}e^{-i\omega_Q t}.
\end{aligned}
\label{eq:joint-model}
\end{equation}
Omitting the second sum gives the linear-only model.

\begin{figure}[t]
  \centering
  \includegraphics{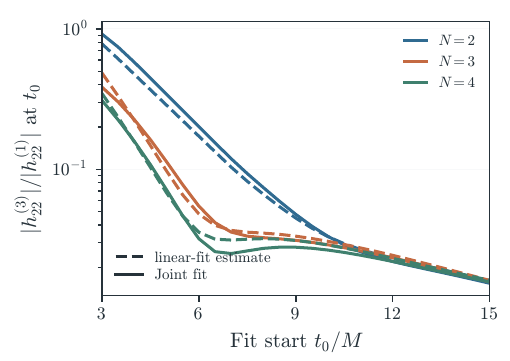}
  \caption{Ratio $|h_{22}^{(3)}|/|h_{22}^{(1)}|$ at each fit start
  $3M\leq t_0\leq15M$. Each sector is summed coherently before taking its
  magnitude. Dashed lines predict the cubic response using parent
  amplitudes from linear-only fits to the full waveform; solid lines use
  amplitudes from joint linear-plus-cubic fits. Both models fit the three
  spherical harmonics simultaneously. Each ratio
  uses the corresponding fit's linear contribution and is evaluated at
  $t=t_0$.}
  \label{fig:start-time-ratio}
\end{figure}

We first fit all three harmonics simultaneously with linear modes alone,
using weighted least squares. Substituting the fitted $I_0$ and $I_1$
amplitudes into Eq.~\eqref{eq:cubic-ratio} predicts the cubic response
without fitting cubic amplitudes. The separately computed
$Z^Q_{\ell 2}$ fix the harmonic projections and are tabulated at
$\chi_f=0.692$ in Appendix~\ref{sec:supp-couplings}.
The dashed lines in Fig.~\ref{fig:start-time-ratio} show the predicted
ratio $|h_{22}^{(3)}|/|h_{22}^{(1)}|$, evaluated at each fit start.
For $N=4$, the predicted cubic-to-total ratio
$|h_{22}^{(3)}|/|h_{22}^{(1)}+h_{22}^{(3)}|$ is $\sim0.37$ at $t_0=3M$,
suggesting an appreciable cubic contribution.
Linear overtones can absorb cubic contributions with similar
frequencies, biasing the parent amplitudes inferred by linear-only fits.
We therefore refit the three harmonics with the linear and predicted
cubic contributions included simultaneously. Cubic amplitudes are
fixed by the linear parents, adding no independent parameters to the
linear-only fit.
Only $A_{I_0}$ and $A_{I_1}$ enter nonlinearly; variable projection
determines the remaining amplitudes by complex weighted least squares. Both models use the same weighted-fitting prescription,
described in Appendix~\ref{sec:supp-fitting}.
The solid lines in Fig.~\ref{fig:start-time-ratio} show the resulting
cubic-to-linear ratios. At $t_0=3M$, the $N=4$ cubic-to-total ratio is
$|h_{22}^{(3)}|/|h_{22}^{(1)}+h_{22}^{(3)}|\simeq0.38$;
At $8M$, the $N=4$ cubic-to-total ratio declines to about $0.03$.

\begin{figure}[t]
  \centering
  \includegraphics{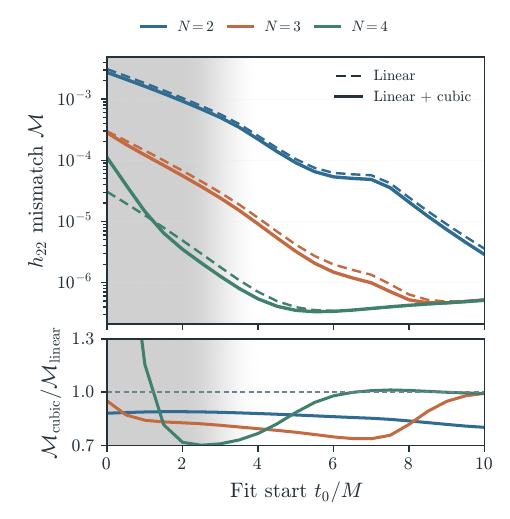}
  \caption{$h_{22}$ mismatch for linear-only (dashed) and linear-plus-cubic
  (solid) fits at each $N$, with equal numbers of fitted amplitudes. The lower
  panel shows the cubic-to-linear mismatch ratio; values below one favor the
  cubic model. The shaded region marks early fit starts where the
  freely decaying linear-plus-cubic QNM model does not yield stable
  linear amplitudes.}
  \label{fig:mismatch}
\end{figure}

To assess whether this sizable contribution improves waveform agreement,
Fig.~\ref{fig:mismatch} compares the $h_{22}$ mismatches of the
linear-only and linear-plus-cubic models, with equal numbers of fitted
amplitudes. Nearby linear overtones can absorb part of the cubic response,
so its magnitude alone does not determine the improvement in mismatch.
Including the predicted cubic terms reduces the mismatch over a range
of fit starts; for $N=4$, the reduction is $29.0\%$ at $t_0=3M$.
This improvement without additional free parameters supports the presence
of cubic response in the waveform. At later starts, the two models approach
a common mismatch floor of order $10^{-6}$, consistent with numerical
noise. As the cubic response decays, this floor limits the resolvable
mismatch improvement.
At the earliest fit starts, the mismatch improvement over linear-only
fits deteriorates, while the inferred amplitudes also drift. Both indicate
limitations of the truncated, freely decaying cubic QNM model in this regime.

\begin{figure}[t]
  \centering
  \includegraphics{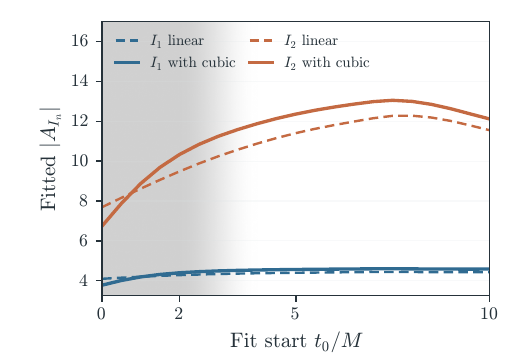}
  \caption{Magnitudes of fitted spherical $I_1$ and $I_2$ amplitudes in
  linear-only (dashed) and linear-plus-cubic (solid) $N=4$ fits as the fit
  start varies. Shading indicates the early-start region with
  pronounced amplitude drift.}
  \label{fig:amplitude-corrections}
\end{figure}

To examine how accounting for cubic response changes the inferred linear
excitation, Fig.~\ref{fig:amplitude-corrections} shows the fitted spherical
$I_1$ and $I_2$ amplitude magnitudes across independent $N=4$ fits with different start times.
At $t_0=3M$, the relative complex-amplitude shifts,
$|A_q^{\mathrm{joint}}-A_q^{\mathrm{linear}}|/|A_q^{\mathrm{linear}}|$,
are $5.3\%$ and $12.8\%$, respectively. These include phase changes and
show that an accurate linear-only waveform fit can yield significantly
biased amplitudes by absorbing nonlinear response, obscuring the
underlying physics.
At the earliest starts, even the truncated linear-plus-cubic model
fails to recover stable linear amplitudes, despite accurately fitting
the total waveform.

These results point to a limitation of the perturbative freely oscillating QNM model. Dynamically excited QNMs with time-dependent excitation amplitudes are expected in this
regime~\cite{DeAmicisEtAl2026Dynamical,DeAmicisEtAl2026DynamicalII,SuEtAl2026Kerr}, and higher-order nonlinearities may also contribute. A full understanding of the early signal components is still lacking.

\paragraph{Discussion.---}
We have developed and implemented a black-hole perturbation framework through third order, obtaining the first quantitative predictions for cubic QNM amplitudes in Kerr spacetime. Using the predicted coupling coefficients, we incorporate cubic modes into joint fits of the $(2,2)$, $(3,2)$, and $(4,2)$ strain multipoles of a numerical-relativity waveform without introducing additional fitting parameters. These fits provide evidence for cubic modes and indicate that their contribution to the dominant $(2,2)$ multipole is substantial during early ringdown.

Our numerical application is limited to one nonprecessing simulation
and six cubic products of the $(2,2)$ fundamental mode, its first overtone,
and their conjugates. The quantitative component sizes, amplitude shifts,
and onset of stability must therefore be tested across other binary
parameters and parent families.

Several directions merit further investigation. First, some cubic modes have oscillation frequencies and damping rates close to those of linear modes. This proximity motivates investigating possible resonant interactions and avoided-crossing behavior, as well as their implications for the excitation of both linear and cubic modes. Second, assessing the detectability of cubic contributions with ground-based and space-based gravitational-wave detectors would establish their observational relevance.

Finally, these coupling coefficients apply when the parent linear QNMs undergo free decay. Early-ringdown excitation may still reflect the evolving driving source. Extending the framework to couplings among dynamically excited QNMs with time-dependent amplitudes would connect cubic response to this source-driven excitation and could improve early-ringdown modeling~\cite{Bucciotti:2026jhr}.

\acknowledgments
\vspace{0.2cm}
\textit{Acknowledgments}---We thank Yanbei Chen for insightful discussions. HY is supported by the Natural Science Foundation
of China (Grant 12573048). NK is supported by the Shuimu
fellowship of Tsinghua University. This work used the \texttt{qnm} Python package~\cite{Stein:2019mop}.
We used OpenAI Codex for assistance with code. The authors reviewed and verified the results.

\appendix
\setcounter{secnumdepth}{1}
\input{prl_supplement.tex}

\bibliography{references}

\end{document}

%% file: flowchart.tikz
  \begin{tikzpicture}[
    x=1cm,y=0.8cm,
    field/.style={draw=flowblue,rectangle,rounded corners=3pt,minimum height=0.725cm,
      minimum width=2cm,inner sep=5pt,fill=flowfill},
    flow/.style={draw=flowink,-{Stealth[length=1.8mm]},line width=0.45pt,
      rounded corners=3pt},
    link/.style={draw=flowink,line width=0.45pt,rounded corners=3pt},
    solve/.style={anchor=south,inner sep=0pt,text height=1.7ex,text depth=0.4ex},
    every node/.style={font=\small}
  ]
        \node[field] (psi0) at (2,5.7) {$\Psi_0^{(k)}$};
    \node[field] (psi4) at (12,5.7) {$\Psi_4^{(k)}$};
    \node[field] (outer) at (2,3.7) {$h_{ab,\mathrm{out}}^{(k)}$};
    \node[field,minimum width=1.4cm] (gauge) at (7,3.7) {$\xi^a_{(k)}$};
    \node[field] (inner) at (12,3.7) {$h_{ab,\mathrm{in}}^{(k)}$};
    \node[field,minimum width=3cm] (jump0) at (2,2)
      {$\Delta\Psi_0^{(k+1)},\ \Delta\partial_R\Psi_0^{(k+1)}$};
    \node[field,minimum width=3cm] (jump4) at (12,2)
      {$\Delta\Psi_4^{(k+1)},\ \Delta\partial_R\Psi_4^{(k+1)}$};
    \node[field,minimum width=4cm] (source0) at (2,0)
      {$S_0^{(k+1)}[h_{ab}^{(k)},\ldots,h_{ab}^{(1)}]$};
    \node[field,minimum width=4cm] (source4) at (12,0)
      {$S_4^{(k+1)}[h_{ab}^{(k)},\ldots,h_{ab}^{(1)}]$};

        \draw[link,dashed,{Stealth[length=1.8mm]}-{Stealth[length=1.8mm]}]
      (psi0.east) -- (psi4.west)
      node[midway,above=1pt,inner sep=1pt]
      {TS identities for $k=1$};
    \draw[flow] (psi0.south) -- (outer.north);
    \draw[flow] (psi4.south) -- (inner.north);
        \node[left=1pt,inner sep=1pt,
      text height=1.5ex,text depth=0.3ex] at (2,4.7)
      {Reconstruction};
    \node[right=1pt,inner sep=1pt,
      text height=1.5ex,text depth=0.3ex] at (12,4.7)
      {Reconstruction};
    \draw[link] (12,4.9) -- (2,4.9);
    \node[above=1pt,inner sep=1pt] at (7,4.9)
      {Boundary conditions at $\mathscr{I}^{+}$};
    \draw[flow] (outer.east) -- (gauge.west)
      node[midway,below=1pt,inner sep=1pt]
      {OLG$\to$ILG};
    \draw[flow] (gauge.east) -- (inner.west)
      node[midway,below=1pt,inner sep=1pt]
      {Boundary conditions};
    \draw[link] (outer.east) -- (3.4,3.7);
    \draw[link] (3.4,3.7) -- (3.4,4.5)
      -- (10.6,4.5) -- (10.6,3.7);

        \draw[link] (gauge.south) -- (7,2.95);
    \draw[link] (7,2.95) -- (2,2.95);
    \draw[link] (7,2.95) -- (12,2.95);
    \node[below=1pt,inner sep=1pt] at (7,2.95)
      {Jump conditions};

        \draw[flow] (outer.south) -- (jump0.north);
    \draw[flow] (inner.south) -- (jump4.north);

        \draw[link] (outer.west) -- (0.2,3.7) -- (0.2,1.05)
      -- (13.8,1.05) -- (13.8,3.7) -- (inner.east);
    \draw[flow] (2,1.05) -- (source0.north);
    \draw[flow] (12,1.05) -- (source4.north);
    \node[below=1pt,inner sep=1pt] at (7,1.05)
      {Reconstructed metric};

        \draw[flow] (source0.west) -- (-0.8,0) -- (-0.8,5.7)
      -- (psi0.west) node[midway,above=1pt,inner sep=1pt] {$k+1\to k$};
    \draw[link,preaction={draw=white,line width=2.4pt}]
      (jump0.west) -- (-0.8,2);
    \draw[flow,draw] (source4.east) -- (14.8,0) -- (14.8,5.7)
      -- (psi4.east) node[midway,above=1pt,inner sep=1pt] {$k\gets k+1$};
    \draw[link,preaction={draw=white,line width=2.4pt}]
      (jump4.east) -- (14.8,2);
        \node[solve,rotate=90] at (-0.88,2.85)
      {Solve $\mathcal{T}_0\Psi_0^{(k+1)}=S_0^{(k+1)}$};
    \node[solve,rotate=-90] at (14.88,2.85)
      {Solve $\mathcal{T}_4\Psi_4^{(k+1)}=S_4^{(k+1)}$};
        \foreach \point in {(2,4.9),(12,4.9),(3.4,3.7),(10.6,3.7),(7,2.95),
      (2,2.95),(12,2.95),
      (2,1.05),(12,1.05),(-0.8,2),(14.8,2)}
      \fill[flowink] \point circle[radius=0.8pt];
  \end{tikzpicture}

%% file: prl_supplement.tex
\section{Newman--Penrose transport equations}
\label{sec:supp-transport}
We reconstruct the metric at each sourced perturbative order using the outer
Newman--Penrose transport system in GHP notation~\cite{NewmanPenrose1962,GerochHeldPenrose1973}.
All perturbative quantities carry
order $k$, which we suppress: these include $h_{ab}$, the reduced Weyl fields
$\Psi_i$, and the spin-coefficient perturbations $\sigma$, $\delta\rho$,
$\delta\tau$ and their barred partners. The Ricci quantities $\Phi_{ij}$ and
$\Lambda$ are computed at the same order from the effective stress-energy tensor.
The GHP operators and unperturbed spin coefficients are evaluated on Kerr;
$\Psi_2^{(0)}$ denotes the background Weyl scalar. Metric subscripts denote
tetrad projections, for example $h_{mm}=h_{ab}m^am^b$ and
$h_{nm}=h_{ab}n^am^b$.

Each transport equation leaves a homogeneous contribution undetermined. For an equation of the form

\begin{equation*}
 (\Th-n_1\rho-n_2\bar\rho)X=S,
\end{equation*}
this contribution decays as $r^{-(n_1+n_2)}$ at null infinity. For some fields, asymptotic flatness requires stronger decay, fixing the homogeneous coefficient to zero. For the remaining fields, complementary Newman--Penrose equations supply the required boundary data.

On complex radial contours, barred partner fields must be continued independently: conjugation relates values on reflected contours and does not supply the barred field on the original contour. Their transport equations follow by interchanging barred and unbarred fields and coefficients, without conjugating the frequency.

Starting from $\Psi_0$, we first reconstruct the shear and angular metric component:

\begin{align}
(\Th-\rho-\bar\rho)\sigma&=\Psi_0,\\
 (\Th+\rho-\bar\rho)h_{mm}&=-2\sigma.
\end{align}
The homogeneous contribution to $\sigma$ decays as $r^{-2}$, consistent with its asymptotic falloff, and therefore requires boundary data. These are determined from $\Psi_4$ through the asymptotic relations below. The homogeneous contribution to $h_{mm}$ approaches a constant; requiring the metric perturbation to decay excludes this term.

We next reconstruct the expansion perturbation and angular trace from the effective Ricci source:

\begin{align}
(\Th-2\rho)\delta\rho&=\Phi_{00},\\
 (\Th+\rho-\bar\rho)h_{m\bar m}&=-2\delta\rho,\\
 \delta\bar\rho&=\delta\rho+h_{m\bar m}(\rho-\bar\rho).
\end{align}
The $r^{-2}$ homogeneous contribution to $\delta\rho$ is compatible with asymptotic flatness. We fix it using residual gauge freedom, choosing the $r^{-2}$ coefficient of $\delta\rho$ to vanish, following~\cite{Price:2006ke}. Requiring $h_{m\bar m}$ to decay then removes its constant homogeneous contribution.

With these fields determined, we reconstruct $\Psi_1$, $\delta\tau$, and $h_{nm}$:

\begin{align}
(\Th-4\rho)\Psi_1&=(\eth'-\tau')\Psi_0\nonumber\\
 &\quad+(\Th-2\bar\rho)\Phi_{01}
 -(\eth-\bar\tau')\Phi_{00},\\
 \Th\delta\tau&=\Psi_1+\Phi_{01}+(\tau-\bar\tau')\delta\rho\nonumber\\
 &\quad+(\bar\tau-\tau')\sigma
 +\tfrac12\rho h_{mm}(\bar\tau-\tau')\nonumber\\
 &\quad+\tfrac12h_{m\bar m}(\tau\bar\rho-\rho\bar\tau'),\\
 (\Th-\bar\rho)h_{nm}&=-2\delta\tau-h_{mm}\tau'
 -h_{m\bar m}\bar\tau'.
\end{align}
The homogeneous contributions to $\Psi_1$ and $h_{nm}$ decay as $r^{-4}$ and $r^{-1}$, respectively, and require the boundary data given below. The homogeneous contribution to $\delta\tau$ approaches a constant and is excluded by asymptotic flatness. Their barred partners are reconstructed in the same way.

Finally, we reconstruct $\Psi_2$ and the remaining radial metric component:

\begin{align}
(\Th-3\rho)\Psi_2&=(\eth'-2\tau')\Psi_1
 +3\delta\rho\Psi_2^{(0)}\nonumber\\
 &\quad-(\Th'-\bar\rho')\Phi_{00}
 +(\eth'-2\bar\tau)\Phi_{01}\nonumber\\
 &\quad-2\Th\Lambda-2\tau\Phi_{10}+2\rho\Phi_{11}.
\end{align}
\begin{align}
-\tfrac12\Th^2h_{nn}={}&\Psi_2+\bar\Psi_2+2\Phi_{11}-2\Lambda\nonumber\\
 &-(\bar\tau-2\tau')(\Th+\rho)h_{nm}\nonumber\\
 &-(\tau-2\bar\tau')(\Th+\bar\rho)h_{n\bar m}\nonumber\\
 &-h_{nm}\bar\rho(\bar\tau+\tau')
 -h_{n\bar m}\rho(\tau+\bar\tau')\nonumber\\
 &-h_{mm}(-\bar\tau^2+\bar\tau\tau'-\tau'^2)\nonumber\\
 &-h_{\bar m\bar m}(-\tau^2+\tau\bar\tau'-\bar\tau'^2)\nonumber\\
 &-h_{m\bar m}\bigl[\tau(-2\bar\tau+\tau')
 +(\bar\tau-2\tau')\bar\tau'\bigr].
\end{align}
The $r^{-3}$ homogeneous contribution to $\Psi_2$ requires boundary data, as does its barred partner. For $h_{nn}$, requiring a decaying solution excludes both the constant and growing homogeneous contributions.

The boundary data follow from $\Psi_4$ through the asymptotic Bianchi identities and Newman--Penrose equations. A superscript ${}^\circ$ denotes the coefficient of a field's leading term in its $1/r$ expansion at null infinity; $\eth^\circ$ and $\eth'{}^\circ$ denote the corresponding asymptotic angular operators. For a mode with $\omega\ne0$,

\begin{align}
\sigma^\circ&=\frac{\bar\Psi_4^\circ}{\omega^2},\\
 \bar\sigma^\circ&=\frac{\Psi_4^\circ}{\omega^2},\\
 \Psi_3^\circ&=\ii\omega\eth^\circ\bar\sigma^\circ,\\
 \bar\Psi_3^\circ&=\ii\omega\eth'{}^\circ\sigma^\circ,\\
 \Psi_2^\circ&=\frac{\eth^\circ\Psi_3^\circ+\Phi_{22}^\circ}
 {-\ii\omega}-2\Lambda^\circ,\\
 \bar\Psi_2^\circ&=\frac{\eth'{}^\circ\bar\Psi_3^\circ+\Phi_{22}^\circ}
 {-\ii\omega}-2\Lambda^\circ,\\
 \Psi_1^\circ&=\frac{\eth^\circ(\Psi_2^\circ+2\Lambda^\circ)
 +2\Phi_{12}^\circ}{-\ii\omega}+\Phi_{01}^\circ,\\
 \bar\Psi_1^\circ&=\frac{\eth'{}^\circ(\bar\Psi_2^\circ+2\Lambda^\circ)
 +2\Phi_{21}^\circ}{-\ii\omega}+\Phi_{10}^\circ,\\
 h_{nm}^\circ&=\eth'{}^\circ\sigma^\circ,\\
 h_{n\bar m}^\circ&=\eth^\circ\bar\sigma^\circ.
\end{align}
These relations determine all required asymptotic coefficients before the radial integrations.

The inner transport system follows by GHP priming. Its boundary data are supplied by the gauge-transformed outer metric at the matching interface.

\section{Complex radial contours and outgoing boundary conditions}
\label{sec:supp-contour}

Imposing outgoing behavior at null infinity requires setting the coefficient
of an exponentially subdominant incoming solution to zero. This is poorly
conditioned on the real radial domain, especially for purely imaginary frequencies,
so we continue the radial equations into the complex plane. Nonlinear products
require all fields on a common contour. Since conjugation reflects that contour
across the real axis, we compute the fields on a pair of reflected contours.

\begin{figure}[htbp]
 \centering
 \includegraphics[width=\linewidth]{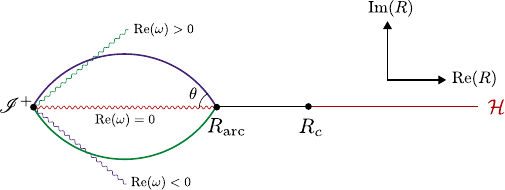}
 \caption{Complex radial contours in $R=1/r$. The smooth arcs join
$\scri^+$ at $R=0$ to $R_{\mathrm{arc}}$ on the real domain;
$R_c=1/r_c$ is the gauge interface, and $\mathcal H$ marks the horizon.
The wavy lines show Stokes directions for the indicated sign of
$\operatorname{Re}\omega$. The arc angle is $\theta=\pi/3$.}\label{fig:supp-contour}
 \label{fig:contour}
\end{figure}

Figure~\ref{fig:contour} shows the two arcs connecting $R_{\mathrm{arc}}$ to $R=0$, above and below the real axis. The coefficient of the subdominant term in the asymptotic representation
changes across a Stokes line, although the solution remains analytic. The outgoing
condition must therefore be imposed in the same Stokes sector as the real axis:
on the lower arc for $\operatorname{Re}\omega>0$ and the upper arc for
$\operatorname{Re}\omega<0$. We obtain the field on the reflected arc by solving a boundary-value problem
with both endpoint values fixed by the first solution. On this arc, the asymptotic
representation contains both incoming and outgoing waves, consistent with the
change in the asymptotic coefficients across the Stokes line.

The arc angle $\theta$ balances convergence on the two contours. For
$\operatorname{Re}\omega>0$, increasing $\theta$ moves the lower arc farther
from the Stokes line, improving convergence. The reflected arc approaches or
crosses the Stokes line, where the incoming wave decays more slowly and oscillates
rapidly, making it harder to resolve. We use $\theta=\pi/3$ to balance the two effects.

For purely imaginary frequencies, the real axis is a Stokes line. For nonlinear
QNMs, these modes generically have $m=0$. We average the prescriptions from the two
arcs to preserve conjugation symmetry. We impose outgoing behavior on each arc in
turn, continue each solution to the opposite arc, and average the two solutions on
both contours.

\section{Cubic coupling coefficients at the remnant spin}
\label{sec:supp-couplings}
For $Q=q_iq_j\bar q_k$, the channel-summed amplitude is
\begin{equation}
 A^{\langle Q\rangle}_{\ell m}
 =Z^Q_{\ell m}A_{q_i}A_{q_j}\bar A_{q_k}.
\end{equation}
The amplitudes use the Letter's $e^{-\ii\omega t}$ convention, with
$t=0$ at the peak of $|h_{22}|$. Our angular convention is
\begin{align}
 S_q(\theta,\phi)&={}_{-2}S_{\ell_qm_q}(\theta;-a\omega_q)e^{\ii m_q\phi}
 \nonumber\\
 &=\sum_{\ell^\prime}C^q_{\ell^\prime}\,
 {}_{-2}Y_{\ell^\prime m_q}(\theta,\phi),\\
 A_q&=C^q_{\ell_q}\mathcal A_q^S,
\end{align}
where $\mathcal A_q^S$ is the spheroidal strain amplitude. Thus the parent
amplitudes are normalized to their corresponding spherical projections.
The table and waveform fits use the same overall strain normalization.

Table~\ref{tab:supp-couplings} gives the six complex coupling coefficients
used in the Letter's waveform fits at the SXS:BBH:0305 remnant spin,
$\chi_f=0.692085$.

Each coefficient includes the channel sum over the nonzero distinct parent modes that contribute to $\omega_Q$.

\begin{table}[htbp]
\centering
\caption{Dimensionless cubic coupling coefficients at $\chi_f=0.692085$ for the six
products of $I_0$, $I_1$, and their conjugates used in the Letter.
The columns give the spherical $(2,2)$, $(3,2)$, and $(4,2)$ projections.}
\label{tab:supp-couplings}
\begin{ruledtabular}
\footnotesize
\setlength{\tabcolsep}{3pt}
\renewcommand{\arraystretch}{1.25}
\begin{tabular}{@{}lccc@{}}
$Q$ & $M_f^2 Z^Q_{22}$ & $M_f^2 Z^Q_{32}$ & $M_f^2 Z^Q_{42}$ \vspace{1em}\\
$I_0^2\bar I_0$ & \shortstack[r]{$0.233595$\\$+0.065330\,\mathrm{i}$} & \shortstack[r]{$0.051180$\\$+0.011079\,\mathrm{i}$} & \shortstack[r]{$0.005156$\\$+0.002817\,\mathrm{i}$} \\[3pt]
$I_0^2\bar I_1$ & \shortstack[r]{$0.096999$\\$+0.039882\,\mathrm{i}$} & \shortstack[r]{$0.015839$\\$-0.007880\,\mathrm{i}$} & \shortstack[r]{$-0.000341$\\$+0.002315\,\mathrm{i}$} \\[3pt]
$I_0I_1\bar I_0$ & \shortstack[r]{$0.340243$\\$+0.023463\,\mathrm{i}$} & \shortstack[r]{$0.067618$\\$-0.014396\,\mathrm{i}$} & \shortstack[r]{$0.005318$\\$+0.000932\,\mathrm{i}$} \\[3pt]
$I_0I_1\bar I_1$ & \shortstack[r]{$0.107680$\\$+0.018397\,\mathrm{i}$} & \shortstack[r]{$0.016792$\\$-0.027034\,\mathrm{i}$} & \shortstack[r]{$-0.002213$\\$+0.005107\,\mathrm{i}$} \\[3pt]
$I_1^2\bar I_0$ & \shortstack[r]{$0.120259$\\$-0.042836\,\mathrm{i}$} & \shortstack[r]{$0.011103$\\$-0.027085\,\mathrm{i}$} & \shortstack[r]{$-0.001928$\\$-0.000490\,\mathrm{i}$} \\[3pt]
$I_1^2\bar I_1$ & \shortstack[r]{$0.032206$\\$-0.014807\,\mathrm{i}$} & \shortstack[r]{$0.008578$\\$-0.015086\,\mathrm{i}$} & \shortstack[r]{$-0.002467$\\$+0.002823\,\mathrm{i}$} \\[3pt]
\end{tabular}
\end{ruledtabular}
\end{table}

\section{Waveform-fitting methodology}
\label{sec:supp-fitting}
Accurate amplitude fits are essential for predicting the cubic response.
In an
exponentially decaying waveform, early-time errors dominate an unweighted
least-squares fit, even when the errors at later times are much smaller.
We therefore use a weighted least-squares fit based on empirical estimates
of systematic and numerical errors, increasing the relative influence of later samples. Additionally,
we use a multimode fit further improve the fits. These
estimates are constructed for each fitting interval and held fixed while
optimizing both waveform models. See Appendix C of Ref.~\cite{Wang:2026rev}
for further discussion.

After obtaining an error estimate $s_{\ell m}(t)$ for the mode $h_{\ell m}(t)$, we minimize
\begin{equation}
 \mathcal L(\{A_q\})=\sum_{t_i\in[t_0,t_f]}\sum_{\ell,m}
 \left|\frac{h^{\rm NR}_{\ell m}(t_i)-h^{\rm model}_{\ell m}(t_i)}
 {s_{\ell m}(t_i)}\right|^2.
\end{equation}

We use the SXS:BBH:0305 waveform at resolution Lev6, with strain extraction Strain\_N2.
For each $t_0$, we perform a simultaneous, unweighted, linear-only
least-squares fit to the three harmonics. The reference basis contains the
spheroidal $(2,2)$, $(3,2)$, and $(4,2)$ families through $n=4$.
Denoting its waveform by $h^{\rm ref}_{\ell m}$, we smooth the absolute
residual and impose a floor:
\begin{align}
 s_{\ell m}(t_i;t_0)&=\max\Bigl[s_{\min},\nonumber\\
 &\quad\underset{|t_j-t_i|\leq\Delta t/2}{\operatorname{mean}}
 \bigl|h^{\rm NR}_{\ell m}(t_j)-h^{\rm ref}_{\ell m}(t_j)\bigr|\Bigr].
\end{align}
The averaging window has width $\Delta t=5M$ and includes only
samples within the fitted interval. This smooths out short-timescale oscillatory
features. We use $s_{\min}=10^{-15}$ in the Letter's waveform normalization.
The reference fit and scales are recomputed for each $t_0$, with the reference
basis, window width, and floor held fixed. For a fixed interval, the same scales
are used for both waveform models and all truncations $N=2,3,4$.

\begin{figure}[htbp]
 \centering
 \includegraphics[width=\linewidth]{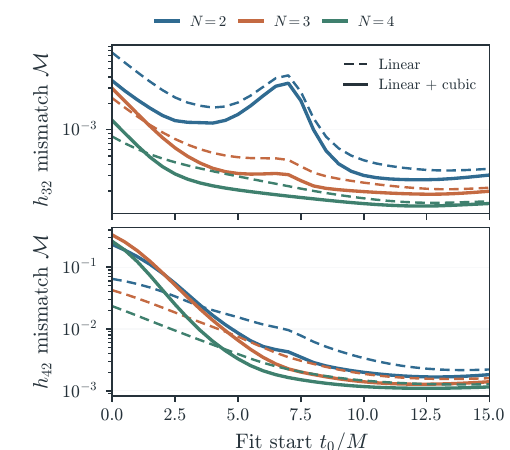}
 \caption{Unweighted $h_{32}$ and $h_{42}$ mismatches for linear-only (dashed) and linear-plus-cubic (solid) models at $N=2,3,4$. Each point comes from a separate fit on $[t_0,60M]$, simultaneous across all three harmonics.}\label{fig:supp-higher-mismatch}
\end{figure}

For the linear-plus-cubic model, only $A_{I_0}$ and $A_{I_1}$ enter
nonlinearly. At each trial value of these parents, we determine the remaining
amplitudes by complex weighted least squares (variable projection). The cubic
amplitudes are fixed by the parent amplitudes and coupling coefficients, so both
models have the same number of fitted amplitudes.

Although the fits are weighted, the per-harmonic mismatch is unweighted:
\begin{align}
 \langle f,g\rangle_{t_0}&=\operatorname{Re}\int_{t_0}^{t_f}f(t)\bar g(t)\,dt,\\
 \mathcal M_{\ell m}(t_0)&=1-
 \frac{\langle h^{\rm NR}_{\ell m},h^{\rm model}_{\ell m}\rangle_{t_0}}
 {\sqrt{\langle h^{\rm NR}_{\ell m},h^{\rm NR}_{\ell m}\rangle_{t_0}
 \langle h^{\rm model}_{\ell m},h^{\rm model}_{\ell m}\rangle_{t_0}}}.
\end{align}

\section{Higher-harmonic waveform agreement}
\label{sec:supp-higher}
The predicted angular structure of the cubic response constrains how it
appears in higher harmonics. Comparing their mismatches therefore tests
whether the improvement extends beyond the dominant harmonic, with the
same parent amplitudes fixing the cubic contributions in each harmonic.
Figure~\ref{fig:supp-higher-mismatch} shows this comparison for $h_{32}$
and $h_{42}$. The cubic model improves $h_{32}$ over nearly all sampled
fit starts, while the improvement in $h_{42}$ depends on fit start and
overtone truncation. At early starts, the $h_{42}$ mismatch can increase when
cubic terms are included. This occurs in the regime where the Letter
also finds drifting fitted amplitudes, indicating the limitations of the
freely decaying QNM model at early times. Once the earliest fit starts are excluded, including cubic modes consistently
reduces the mismatch in both harmonics across all three truncations.

\section{Waveform components and cancellation}
\label{sec:supp-components}
\begin{figure}[htbp]
 \centering
 \includegraphics[width=\linewidth]{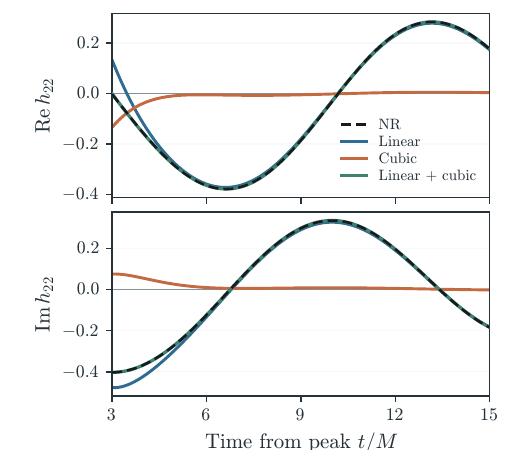}
 \caption{Real and imaginary parts of spherical $h_{22}$ from the fixed $N=4$ joint fit starting at $t_0=3M$, compared with NR. The linear, cubic, and total waveforms use the same fitted amplitudes throughout the displayed interval.}\label{fig:supp-cubic-waveform}
\end{figure}

To relate the fitted cubic amplitudes to their net contribution to the waveform, we show both the linear--cubic decomposition and the individual cubic products.
Figure~\ref{fig:supp-cubic-waveform} compares the real and imaginary parts
of the fitted $h_{22}$ waveform with NR, showing the linear and cubic
contributions separately. The amplitudes are fixed by the $N=4$ joint fit
starting at $t_0=3M$ and are held fixed throughout the displayed interval.
The total waveform follows NR more closely than the linear contribution
alone.

\begin{figure}[htbp]
 \centering
 \includegraphics[width=\linewidth]{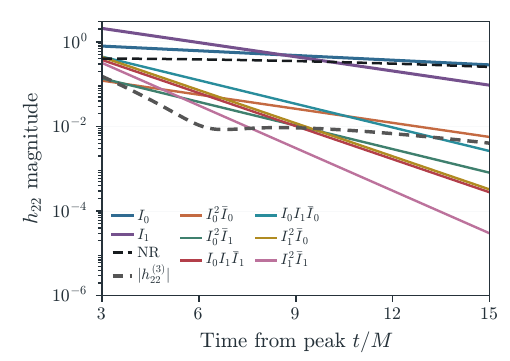}
 \caption{Magnitudes of the linear parents $I_0,I_1$, the six cubic products,
and their coherent sum $|h_{22}^{(3)}|$, compared with NR. The model curves use
the fixed joint fit shown in Fig.~\ref{fig:supp-cubic-waveform}, with the same
spherical $h_{22}$ normalization. Only the two linear parents are displayed;
the fit includes the other linear modes.}\label{fig:supp-mode-cancellation}
\end{figure}

Figure~\ref{fig:supp-mode-cancellation} shows the instantaneous magnitudes
of the six cubic products and their coherent sum for the same fit, with
the linear parents $I_0,I_1$ and NR included for comparison. Several individual cubic contributions exceed
their coherent sum, displaying cancellations analogous to those observed
in linear overtone fits.
